\documentclass[pdflatex,sn-mathphys-num]{sn-jnl}

\usepackage{graphicx}%
\usepackage{multirow}%
\usepackage{amsmath,amssymb,amsfonts}%
\usepackage{amsthm}%
\usepackage{mathrsfs}%
\usepackage[title]{appendix}%
\usepackage{xcolor}%
\usepackage{textcomp}%
\usepackage{manyfoot}%
\usepackage{booktabs}%
\usepackage{algorithm}%
\usepackage{algorithmicx}%
\usepackage{algpseudocode}%
\usepackage{listings}%

\usepackage{algorithm}
\usepackage{algpseudocode}

\theoremstyle{thmstyleone}%
\theoremstyle{thmstyletwo}%

\theoremstyle{thmstylethree}%

\begin{document}

\title[Article Title]{iHAC: A Hybrid Cluster Architecture for Enhanced Performance and Resilience}


\author*[1,4]{\fnm{Siddique Abubakr} \sur{Muntaka}}\email{muntaksr@mail.uc.edu}

\author[2]{\fnm{Edward} \sur{Danso Ansong}}

\author[3]{\fnm{Benjamin} \sur{Yankson}}

\author[4]{\fnm{Oliver} \sur{Kornyo}}

\author[5]{\fnm{Faiza} \sur{Hussein}}

\author[6]{\fnm{Mohammed Nadhir} \sur{Muntaka}}

\author[7]{\fnm{Joshua} \sur{Dagadu}}

\author[7]{\fnm{Prince} \sur{Clement Addo}}

\author[7]{\fnm{Maxwell} \sur{Dorgbefu Jnr.}}

\author[7]{\fnm{Franco} \sur{Osei-Wusu}}

\author[8]{\fnm{Foster} \sur{Yeboah}}

\author[4]{\fnm{Michael} \sur{Asante}}

\affil [1]{\orgdiv{School of Information Technology}, \orgname{University of Cincinnati}, \city{Cincinnati}, \state{Ohio}, \country{USA}}

\affil [2]{\orgdiv{Department of Computer Science}, \orgname{University of Ghana}, \city{Accra}, \country{Ghana}}

\affil[3]{\orgdiv{Department of Information Sciences}, \orgname{University at Albany}, \city{Albany}, \state{New York}, \country{USA}}

\affil[4]{\orgdiv{Department of Computer Science}, \orgname{Kwame Nkrumah University of Science and Technology}, \city{Kumasi}, \state{Ashanti}, \country{Ghana}}

\affil[5]{\orgdiv{Computer Science Department}, \orgname{Garden City University College}, \city{Kumasi}, \state{Ashanti}, \country{Ghana}}

\affil[6]{\orgdiv{Penn State I.T., Learning and Performance Systems \& CIED}, \orgname{Pennsylvania State University}, \city{State College}, \state{PA}, \country{USA}}

\affil[7]{\orgdiv{Department of Information Technology Education}, \orgname{Akenten Appiah-Menka University of Skills Training and Entrepreneurial   Development. (AAMUSTED)}, \city{Kumasi}, \state{Ashanti}, \country{Ghana}}

\affil [8]{\orgdiv{College of Engineering and Applied Science}, \orgname{University of Cincinnati}, \city{Cincinnati}, \state{Ohio}, \country{USA}}


\abstract{Uninterrupted system availability is a critical requirement for enterprise operations, yet traditional high-availability clusters suffer from limitations like single points of failure and inefficient resource allocation. This paper introduces and evaluates the Integrated High Availability Cluster (iHAC), a hybrid architecture designed to enhance system resilience and performance. The iHAC integrates the strengths of active-active and active-passive configurations to optimize workload distribution and failover capabilities. We conducted a comparative analysis, simulating iHAC against conventional (legacy) clusters using Riverbed Modeler (OPNET). The results reveal significant performance improvements: iHAC reduced the average HTTP page response time by over 40\%, from five seconds in a traditional active-active setup to under three seconds. This was achieved alongside diminished network latency and increased overall throughput. This study validates the iHAC architecture as a superior design for building robust, high-performance systems, offering a practical path to greater operational continuity and resilience.}




\maketitle

\section{Introduction}\label{sec1}
\begin{figure}[h]
\centering
\includegraphics[width=0.9\textwidth]{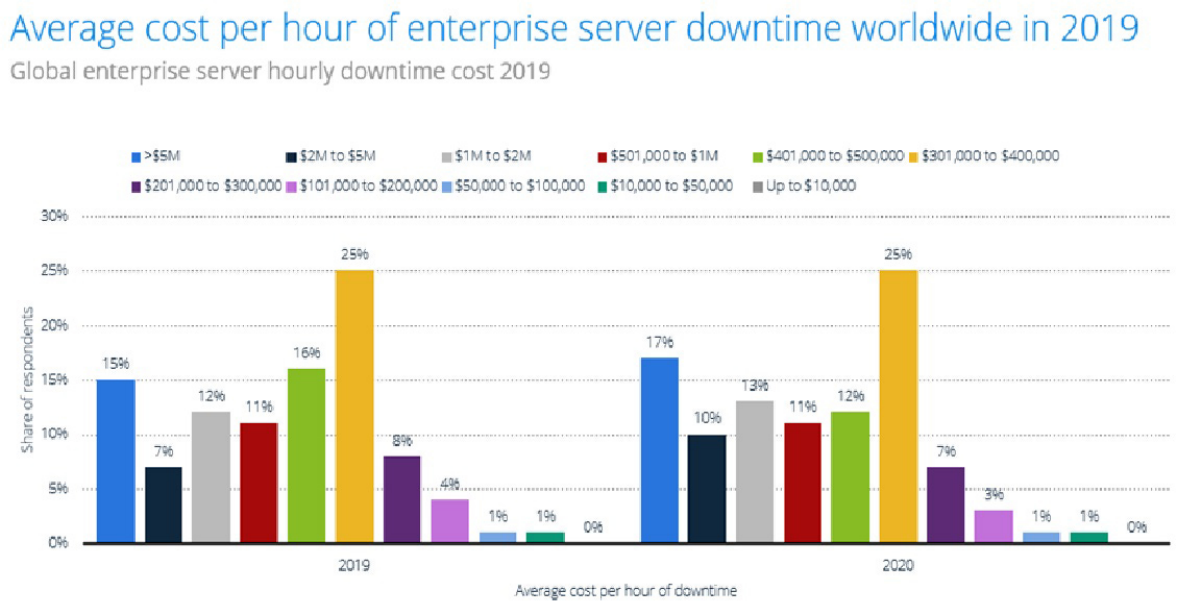}
\caption{Report on the Average Cost/Hour for Server Downtime Worldwide \cite{statista2019downtime}}
\label{Figure:serverdowntime}
\end{figure}

In today's digitally networked environment, server availability is a cornerstone of business continuity and effective service delivery \cite{silva2025models}. Organizations increasingly rely on services that must operate continuously to meet user demands, making the mitigation of system downtime a critical challenge \cite{thaqi2024premises}. Any significant service interruption, whether caused by hardware faults, software bugs, or configuration errors, can lead to severe operational and financial consequences.  The cost of enterprise server downtime can be staggering, as illustrated in Figure \ref{Figure:serverdowntime}. For example, studies have shown that a lot of businesses have lost hundreds and thousands of dollars every hour, with 15\% of them losing more than \$5 million every hour according to studies in \cite{wang2020enterprise}. Thus, the need for strong High-Availability (HA) systems that can guarantee operational continuity is therefore highlighted by these financial consequence \cite{wang2020enterprise}. 

Conventional (legacy) HA solutions typically rely on cluster architectures like active-active and active-passive designs that provides redundancy. While these techniques are widely used, they are traditional architectures or designs that come with inherent limitations. According to Mihai et al., active-active clusters, which distributes workload across all nodes, can be susceptible to cascading failures that disrupt the entire system \cite{mihai2024security}. Conversely, active-passive clusters leave expensive backup resources idle until a failure occurs, leading to inefficient resource utilization \cite{kim2024design}. These deficiencies highlight the need for a more advanced and hybrid approach that can combine the benefits of both architectures while mitigating the respective weaknesses inherent in the legacy systems. 
To address these limitations, this study introduces and evaluates an Integrated High Availability Cluster (iHAC) as a hybrid architecture designed to enhance system resilience and performance. The iHAC architecture integrates the strengths of active-active and active-passive configurations to optimize workload distribution and failover capabilities. The study adopts a computational approach using simulation software called Riverbed Modeler (formerly known as OPNET) to conduct a comparative performance analysis of iHAC against conventional (legacy) cluster architectures. The primary objectives of this research are as follows:

\begin{enumerate}
    \item To design and develop a hybrid high-availability cluster architecture (iHAC) that integrates active-active and active-passive principles.
    \item To implement advanced load-balancing techniques within the iHAC to optimize resource utilization and system responsiveness.
    \item To conduct a rigorous, simulation-based performance evaluation of the iHAC architecture against traditional cluster configurations.
    \item To analyze the effectiveness of the iHAC in enhancing system resilience and fault tolerance based on performance metrics such as response time, throughput, and latency.
\end{enumerate}

The subsequent sections of this paper are structured as follows: Section 2 critically examines the literature on HA systems, identifying key theoretical and practical gaps. Section 3 elaborates on the methodological governing of iHAC’s design and simulation. Section 4 presents the findings and discusses their implications for HA system optimization. Finally, Section 5 consolidates the study’s contributions and proposes avenues for future research in adaptive fault-tolerant architectures.

\section{Background \& Related work}\label{sec2}

High Availability (HA) is a foundational principle in systems design \cite{mesbahi2018reliability}. It aims to ensure continuous operational uptime and resilience against service interruptions \cite{umam2018implementation} \cite{muntaka2025resilience}. In modern enterprise environments, services are expected to be perpetually accessible. HA architectures are therefore essential for mitigating the risks of hardware faults, software bugs, and other system failures \cite{xinming2019research}. The effectiveness of these systems is traditionally quantified using standard reliability metrics. The Mean Time Between Failures (MTBF) as expressed in \cite{zyluk2023implementation} indicates how long a system operates without malfunction, it is expressed as:

\begin{equation}
\text{MTBF} = \frac{\text{Total Operational Time}}{\text{Number of Failures}} \label{eq:mtbf}
\end{equation}

A high MTBF value implies greater system dependability. Żyluk et~al.~\cite{zyluk2023implementation} explained that the Mean Time to Repair (MTTR) quantifies the average time required to restore a system following a failure. It is expressed mathematically as:

\begin{equation}
\text{MTTR} = \frac{\text{Total Downtime}}{\text{Number of Failures}} \label{eq:mttr}
\end{equation}

A low MTTR value signifies a faster recovery process. The overall System Availability (A), a critical measure of a system's operational status, is then derived from these metrics. A higher availability percentage, as calculated using the following mathematical model, indicates a more robust and dependable system:

\begin{equation}
A = \frac{\text{MTBF}}{\text{MTBF} + \text{MTTR}} \label{eq:availability}
\end{equation}

\subsection{Conventional High-Availability Cluster Architectures}\label{subsec2}

To achieve high availability, organizations widely employ server clustering with multiple interconnected servers, or nodes, that function as a single, resilient system \cite{govindan2016evolve}. These clusters are designed to eliminate single points of failure through redundancy. The two most common conventional configurations are the active-passive and active-active designs. The active-passive architecture (design), often referred to as a hot-standby configuration, designates one server as the primary node to handle all client requests while a secondary node remains idle. This passive node continuously monitors the primary's health, typically via a "heartbeat" signal \cite{muntaka2025enhanced}. If the primary node fails, the passive node is automatically promoted to the active state to take over its operations, ensuring service continuity. While this design provides a straightforward and reliable failover mechanism, its principal drawback is the significant underutilization of resources, as expensive backup hardware remains inactive during normal operations \cite{kim2024design}.

\begin{figure}[h]
\centering
\includegraphics[width=0.9\textwidth]{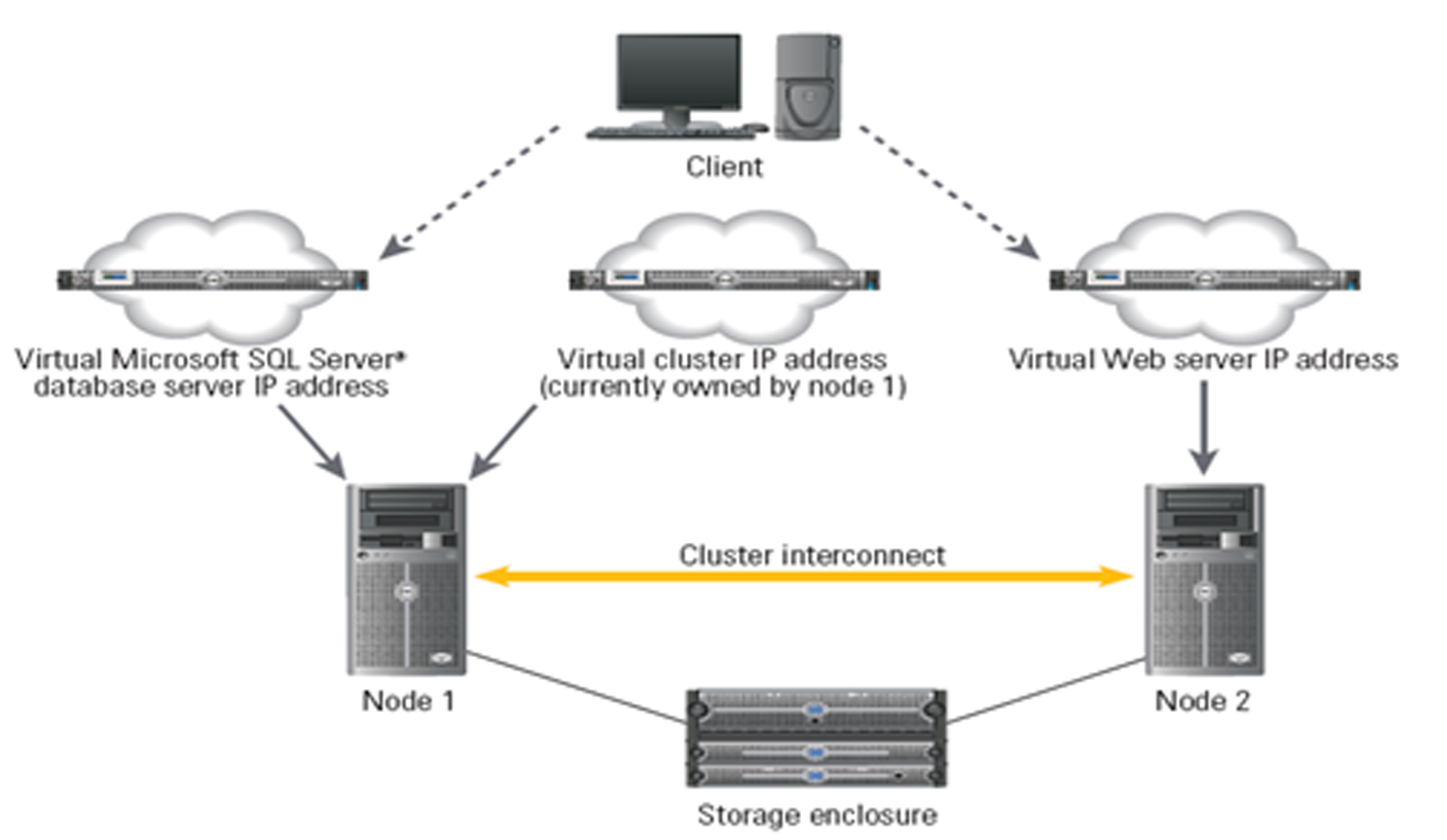}
\caption{Microsoft Cluster Server architecture based on Active/active Cluster design/framework \cite{muntakahybrid}}
\label{Figure:MS_activeactive}
\end{figure}

In contrast, the active-active architecture utilizes all nodes in the cluster simultaneously to serve client requests, typically employing a load balancer to distribute the workload. This approach maximizes resource utilization and can offer superior performance under heavy loads \cite{mihai2024security}. However, active-active configurations are inherently more complex to manage, requiring sophisticated load balancing and data synchronization mechanisms to maintain a consistent state across all nodes. Furthermore, they can be more susceptible to cascading failures, where a fault on one node can potentially impact the stability of the entire cluster \cite{engelmann2008symmetric}. A related concept, the N+1 architecture, offers a compromise by having N active nodes supported by a single standby node, balancing redundancy and cost \cite{rick2023scalability}. A prime example of these principles in practice is the Microsoft Cluster Service (MSCS), which facilitates the failover of services between nodes in a cluster to maintain availability for clients, as illustrated in Figure \ref{Figure:MS_activeactive}

\subsection{Advanced Paradigms in Modern Environments}\label{subsec2}

The evolution of computing has introduced more advanced paradigms for achieving high availability, particularly within virtualized, cloud, and large-scale distributed systems. \\
\textbf{Virtualization and Cloud Computing:} The proliferation of virtualization has enabled more dynamic HA strategies, allowing for the rapid provisioning and migration of virtual machines to facilitate flexible failover and recovery \cite{toy2017high}. In cloud computing environments, the challenges expand to include efficient resource allocation and job scheduling across multi-tenant infrastructures. Research in this area has explored solutions like the Optimal Physical Host with Load Balancing (OPH-LB) design, which uses probabilistic methods to optimize resource allocation and enhance availability in Infrastructure as a Service (IaaS) platform \cite{chhabra2018probabilistic}. For hybrid environments that combine on-premises and cloud resources, a Weighted Availability (WA) can be calculated to reflect the contributions of each component. It is mathematically expressed as:

\begin{equation}
WA = \alpha \cdot A_{\text{On-Prem}} + \beta \cdot A_{\text{Cloud}} \label{eq:wa}
\end{equation}

\begin{figure}[h]
    \centering
    \includegraphics[width=0.99\linewidth]{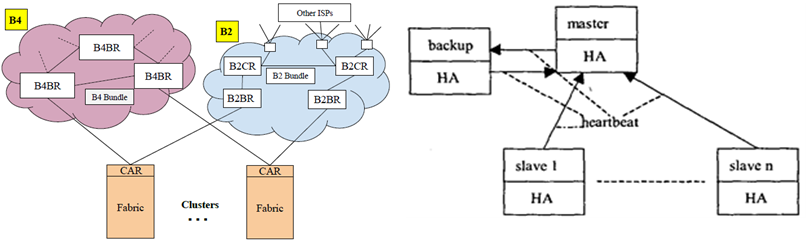}
    \caption{Design of High Availability Cluster in Google Network Network \cite{govindan2016evolve} }
    \label{fig:googleHA}
\end{figure}

Where $\alpha$ and $\beta$ represent the proportional weights of the on-premises and cloud resources, respectively. \\
\textbf{Replication and Fault Tolerance:} Replication is another cornerstone of HA, ensuring that data and services are mirrored across multiple locations. In the context of RESTful web services, for instance, the ReServE platform explores various fault-tolerance techniques beyond simple state-machine replication, such as message logging and request retries, to enhance service reliability \cite{kobusinska2018towards}. These methods provide service providers with the flexibility to choose recovery mechanisms best suited to their needs.\\
\textbf{Large-Scale Industrial Implementations:} The principles of high availability are perhaps best demonstrated in the architectures of large-scale internet companies. Google’s B4 network, for example, employs extensive redundancy across its user-cluster and inter-cluster communications, utilizing heartbeat signals between master and backup nodes for automatic failover \cite{govindan2016evolve}. As depicted in Figure \ref{fig:googleHA}, such designs exemplify the scalability and resilience required to maintain global service operations, providing valuable insights into robust HA design patterns.

\subsection{Identifying the Research Gap}\label{subsec2}

The existing literature establishes a fundamental dilemma in conventional cluster design. The active-passive prioritizes simplicity and failover reliability, but at the significant cost of resource inefficiency. The active-active prioritizes performance and resource utilization, but at the expense of increased complexity and a greater risk of system-wide disruption. While advanced paradigms in cloud and virtualized environments offer greater flexibility, the core trade-off between these two fundamental approaches remains a persistent challenge.
This analysis reveals a distinct research gap: a need for a hybrid cluster architecture that can strategically combine the strengths of both canonical designs. Such a solution must aim to balance high performance and efficient resource utilization with simple, robust, and reliable failovers. By doing so, it can offer a more holistic and practical approach to high availability than traditional configurations alone. This paper addresses this specific gap through the design and rigorous performance evaluation of the iHAC model.

\section{Methodology}\label{sec3}

Using design science research, we employed qualitative analysis of simulation-based experiments as supported in the scholarly works of \cite{susilawati2025research}. The methodology uses a model-based approach to comprehensively assess high-availability (HA) cluster systems and their real applications. \cite{elio2011computing} explained that model methodology involves the abstraction of real-world systems to create simplified versions that retain their essential features. The approach is therefore suitable for use and can help research to focus on specific characteristics and evaluate performance in a controlled environment. This avoids the challenges associated with testing on operational or production-based systems, where manual processes for essential tasks like configuration backups can introduce significant risks to operational continuity, as demonstrated in the case of a large hospital network by Osei-Wusu et al. \cite{osei2025automating} The abstraction capabilities can make it easy to identify the crucial parameters that affect a system’s performance. In iHAC, we employ this methodology to clarify HA clustering techniques, thereby enhancing system uptime and resource utilization.

\subsection{Simulation Environment}\label{subsec3}

The study utilizes Riverbed Modeler Academic Edition version 17.5 software for modelling and simulating iHAC. The software is a sophisticated network simulation tool that mimics real-world systems in an experiment and testbed setting. Riverbed has a graphical design of networks with nodes, connections, and subnets \cite{aslam2024review}. This simulation software is therefore said to feature a user-friendly interface with versatility and traceability and improved user experience. This makes it well-suited for modelling complex systems.  The experiment evaluates iHAC in three scenarios, namely, Active-Active Passive (A-AP), Active-Passive Active (A-PA), and Active-Passive Passive (A-PP). Performance metrics used are; response time, throughput, resource use, and recovery duration.

\subsection{iHAC Framework Design}\label{subsec3}

\begin{figure}[h]
\centering
\includegraphics[width=0.99\textwidth]{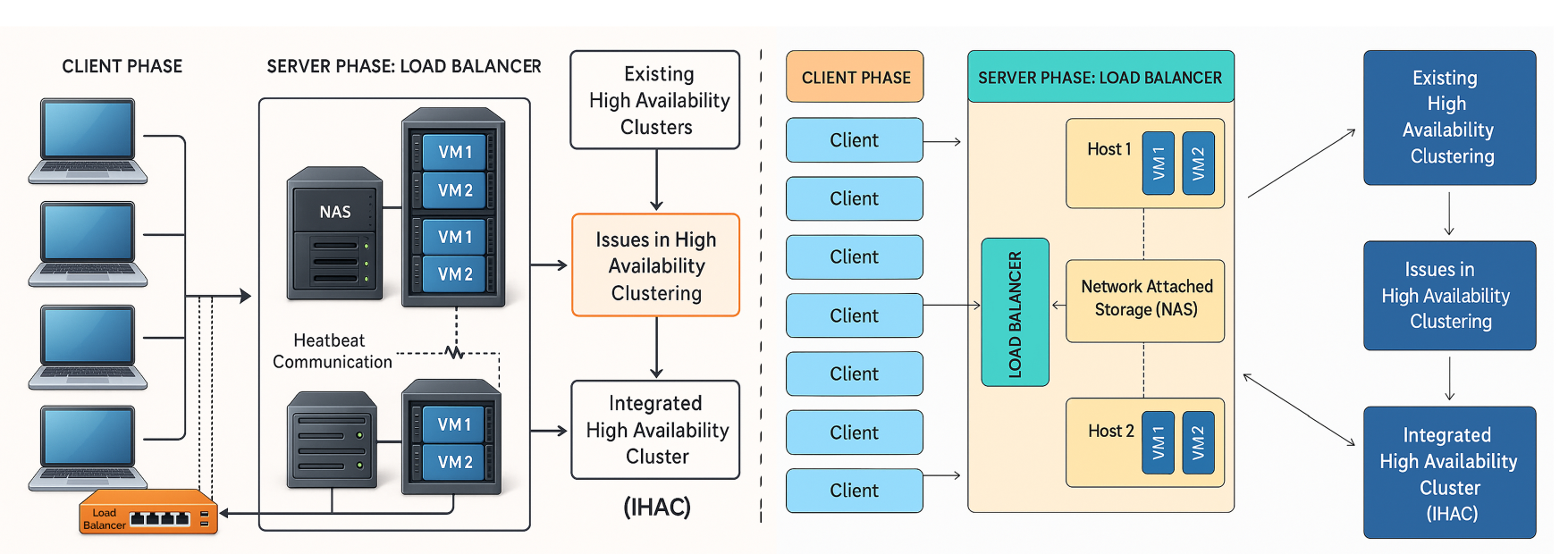}
\caption{Integrated high availability cluster architecture framework (iHAC)}
\label{Figure:Mihacframework}
\end{figure}

As shown in Figure ~\ref{Figure:Mihacframework}, the iHAC system framework is structured into three core components: the \textit{Client Network}, the \textit{Load Balancer}, and the \textit{Server Cluster}. These are organized into two distinct operational phases (client phase and server phase), which separate the layers of interaction.

When an end-user initiates a resource request within the client network, the request is intercepted by the load balancer. This component serves as a dynamic intermediary, distributing client requests across available servers based on integrated load-balancing algorithms. The server cluster, in turn, combines \textit{active-active} and \textit{active-passive} configurations to enhance redundancy and optimize resource utilization. The load-balancing mechanism supports multiple strategies, including \textit{Least Connection}, \textit{Server Load}, and \textit{Random Selection} (LSR), ensuring flexible and efficient traffic management.

To formally define the operational logic of iHAC, Algorithm~\ref{alg:ihac} presents the core request handling and failover process. It illustrates how incoming requests are routed to active nodes while a concurrent heartbeat monitoring system promotes passive nodes in the event of failure.

\begin{algorithm}[H]
\caption{iHAC Request Handling and Failover Logic}
\label{alg:ihac}
\begin{algorithmic}[1]
\Require Incoming Client Request $R$, Set of all Nodes $S$
\Statex \textbf{Data:} Node Status $S.\text{status}$ for each node in $S$ (e.g., ACTIVE, PASSIVE, FAILED)
\While{true} \Comment{Continuous Operation}
    \If{new client request $R$ is received}
        \State \textbf{// Identify currently operational primary servers}
        \State $Active\_Nodes \gets \{ node \in S \mid node.status = \text{ACTIVE} \}$
        \If{$Active\_Nodes \neq \emptyset$}
            \State \textbf{// Use load balancing to select the best active node}
            \State $Target\_Node \gets Load\_Balance(Active\_Nodes, R)$
            \State \textbf{Forward} $R$ \textbf{to} $Target\_Node$
        \Else
            \State \textbf{// No active nodes available, service is temporarily down}
            \State Queue $R$ \textbf{or return} \texttt{503 Service Unavailable}
        \EndIf
    \EndIf
    \ForAll{$node \in S$ \textbf{where} $node.status = \text{ACTIVE}$}
        \If{$Heartbeat\_Missed(node)$}
            \State \textbf{// Failure detected, initiate failover}
            \State $node.status \gets \text{FAILED}$
            \State $Passive\_Nodes \gets \{ p \in S \mid p.status = \text{PASSIVE} \}$
            \If{$Passive\_Nodes \neq \emptyset$}
                \State \textbf{// Promote a passive node to active}
                \State $New\_Active \gets Select\_Best\_Passive(Passive\_Nodes)$
                \State $New\_Active.status \gets \text{ACTIVE}$
            \EndIf
        \EndIf
    \EndFor
\EndWhile
\end{algorithmic}
\end{algorithm}

\subsection{Network Topology}\label{subsec3}

\begin{figure}[h]
    \centering
    \includegraphics[width=0.92\linewidth]{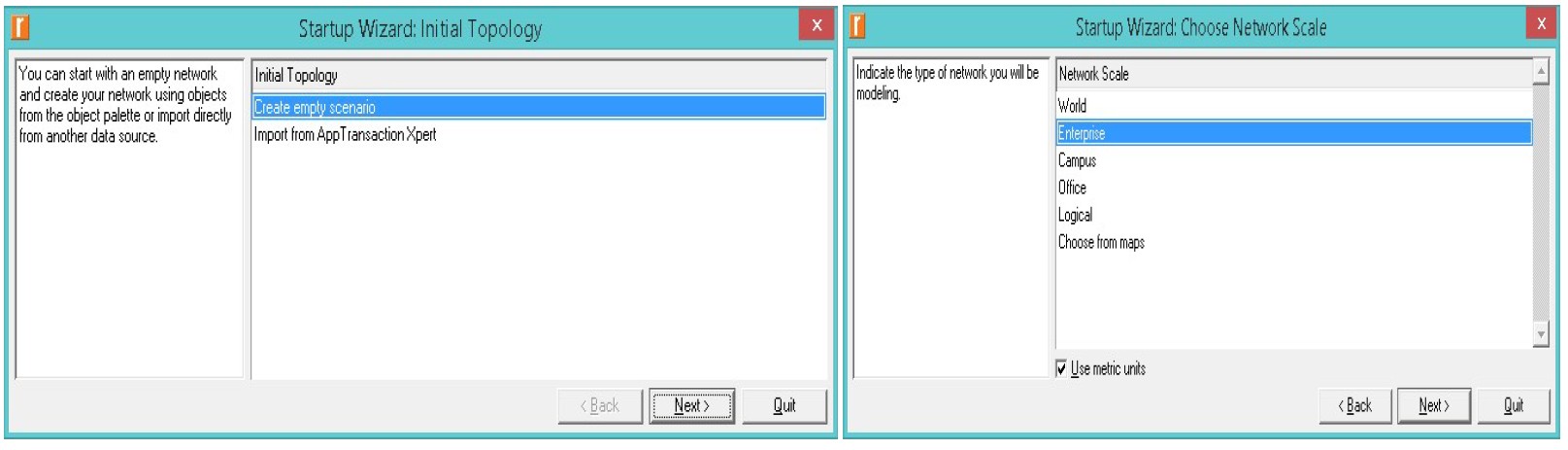}
    \caption{Creating and choosing a network topology in Riverbed Modeler (OPNET)}
    \label{fig:fig5}
\end{figure}

\begin{figure}[h]
    \centering
    \includegraphics[width=1\linewidth]{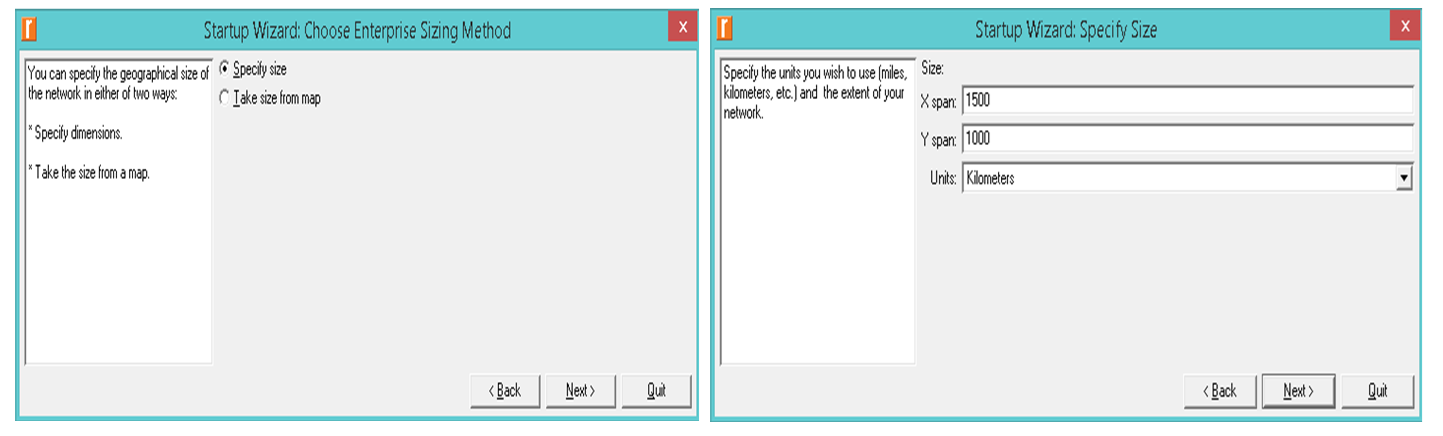}
    \caption{Sizing a network in Riverbed Modeler}
    \label{fig:fig6}
\end{figure}

As found in Figures \ref{fig:fig5} and \ref{fig:fig6}, this helps illustrate the process of creating a network topology using the Riverbed Modeler. The topology is defined as an enterprise-scale network in which the client side is connected through a gateway router. In contrast, on the server side are the load balancers, switches, and clustered servers that run Hypertext Transfer Protocol (HTTP) and other protocols or application suite requests and processing. The network is scaled to accommodate the required components.

\subsection{Experiment and Modelling in Riverbed}\label{subsec3}

\begin{figure}[h]
    \centering
    \includegraphics[width=0.95\linewidth]{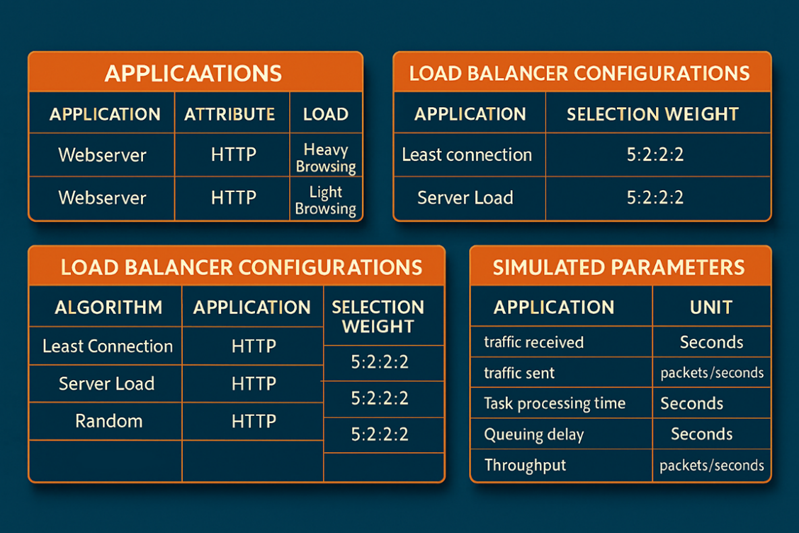}
    \caption{Application Description, Load Balancer Configurations, and Simulated Parameters}
    \label{fig:fig7}
\end{figure}

\begin{figure}[h]
    \centering
    \includegraphics[width=0.95\linewidth]{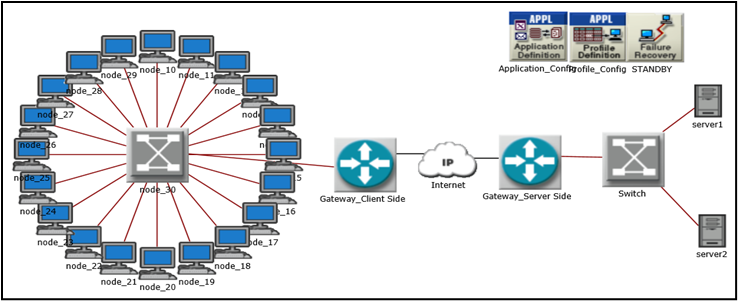}
    \caption{Active-Active Cluster}
    \label{fig:fig8}
\end{figure}

\begin{figure}[h]
    \centering
    \includegraphics[width=0.90\linewidth]{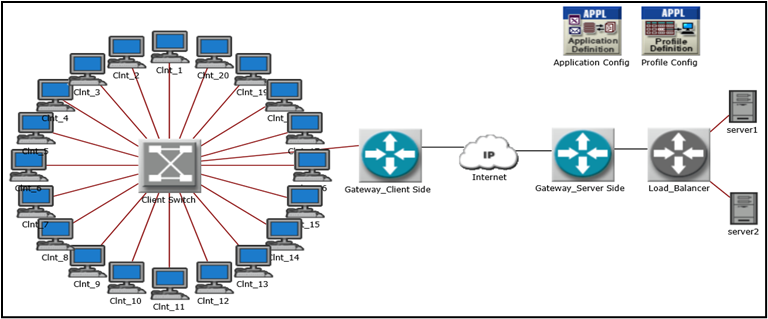}
    \caption{Active-Passive Cluster}
    \label{fig:fig9}
\end{figure}

\begin{figure}[h]
    \centering
    \includegraphics[width=0.90\linewidth]{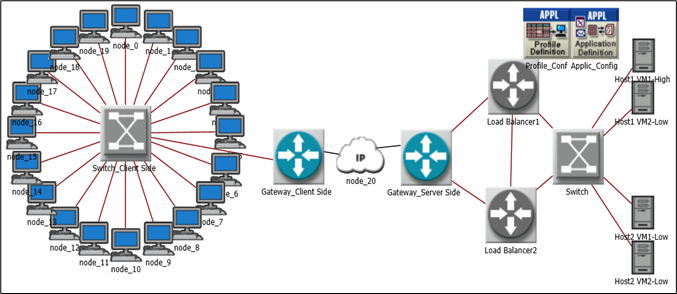}
    \caption{Integrated High Availability Cluster (iHAC)}
    \label{fig:fig10}
\end{figure}

\begin{figure}[h]
    \centering
    \includegraphics[width=0.95\linewidth]{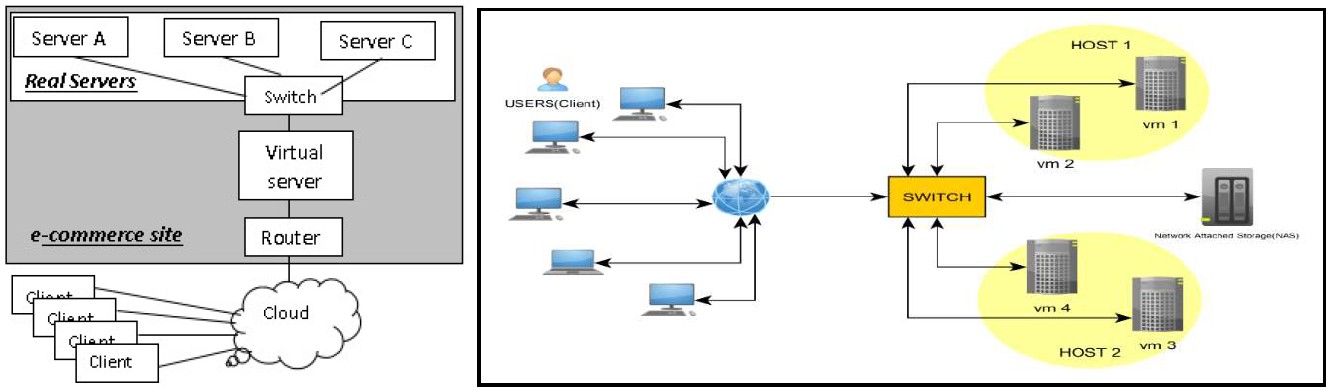}
    \caption{iHAC Architectural Components and Deployment Scenario}
    \label{fig:fig11}
\end{figure}

The system’s attributes, load levels, and load balancer configurations are detailed in Figure \ref{fig:fig7}, specifying selection weights for each algorithm. Key simulation parameters, including response time, throughput, and queuing delay, were evaluated under different scenarios to assess the effectiveness of iHAC. To optimize performance, the iHAC framework leverages advanced load-balancing algorithms such as Least Connection, Server Load, and Random Selection. The Least Connection algorithm directs requests to servers with the fewest active connections, while the Server Load algorithm routes traffic based on real-time CPU and memory usage. The Random Selection algorithm probabilistically allocates requests, enhancing adaptability in unpredictable conditions. These algorithms ensure efficient and resilient client request handling across server clusters. A comparative analysis of traditional and iHAC is illustrated in Figures \ref{fig:fig8}, \ref{fig:fig9}, \ref{fig:fig10}, and \ref{fig:fig11}. Figure \ref{fig:fig8} depicts an Active-Active Cluster, where all servers operate simultaneously to maximize resource utilization. Figure \ref{fig:fig9} presents an Active-Passive Cluster, where only the primary server handles requests while the secondary remains on standby for failover support. Figure \ref{fig:fig10} illustrates the Integrated High Availability Cluster (iHAC), showcasing its hybrid architecture combining active-active and active-passive principles for superior resilience. Figure 11 is divided into left and right sections, demonstrating server balancing mechanisms of the iHAC. The left side features a Content Switch dynamically distributing client requests (received via a router) to servers A, B, or C based on real-time load metrics. The right side highlights the iHAC architecture scenario. Backup servers operate in active and passive modes to ensure rapid recovery with minimal downtime. Additionally, as shown in Figures 10 and 11, the iHAC deployment incorporates virtual switches and load balancers to replicate real-world conditions. After the system processes incoming client requests and applies scheduling algorithms, the virtual switch routes traffic to the most suitable server based on factors such as resource consumption, active connections, and probabilistic distribution. The Active-Passive Active (A-PA) reduces resource contention while marginally prolonging recovery time by prioritizing failover through heartbeat monitoring of the systems. As a redundant backup for hardware reliability, the Active-Passive Passive (A-PP) setup maintains secondary servers in a warm or cold standby without requiring active resource expenditure. By ensuring effective workload distribution, this approach enhances overall system stability and prevents server overload issues.

\subsection{Performance Evaluation Model}\label{subsec3}

Metrics such as availability, resource usage, and failover efficiency were used to evaluate iHAC. In mathematical construct, the ratio of uptime to total system time is measured by availability ($A$). Hence, expressed as:

\begin{equation}
A = \frac{\text{MTBF}}{\text{MTBF} + \text{MTTR}} \label{eq:availability}
\end{equation}

The average operational period without failure is expressed in terms of \textit{mean time between failures} (\textit{MTBF}). \textit{Mean time to repair} (\textit{MTTR}) assesses the average downtime necessary to restore the system. A higher availability ($A$) value indicates superior system reliability, which is an essential prerequisite for continuous operations in enterprise environments.

\textit{Resource Utilization}, represented as ($U$), measures the efficiency of resources allocated among active and inactive servers within a cluster. This is mathematically articulated as:

\begin{equation}
U = \frac{\sum_{i=1}^{n} W_i \cdot R_i}{\sum_{i=1}^{n} R_i} \label{eq:utilization}
\end{equation}

Where $W_i$ denotes the weight allocated to server $i$, and $R_i$ signifies the resources utilized by server $i$. The iHAC optimizes the use of computational resources by dynamically adjusting $W_i$ according to the prevailing system conditions. This reduces waste and enhances system efficiency.

\textit{Weighted Availability} ($WA$) considers the hybrid characteristics of the iHAC system. It integrates the availability measurements of both on-premises and cloud resources by mathematically representing it as:

\begin{equation}
WA = \alpha \cdot A_{\text{On-Prem}} + \beta \cdot A_{\text{Cloud}} \label{eq:wa}
\end{equation}

Where $\alpha$ and $\beta$ denote the weights allocated to on-premises and cloud components, respectively, such that $\alpha + \beta = 1$. The indicators measure the equitable contribution of both components to the system's reliability.

\textit{Failover Efficiency} ($F$) of the iHAC model is a vital indicator that reflects the system's responsiveness during failure-induced transitions. It is delineated as:

\begin{equation}
F = \frac{1}{T_{\text{failover}}} \label{eq:failover}
\end{equation}

$T_{\text{failover}}$ refers to the duration the load balancer requires to reroute traffic to a functional server upon identifying a failure. A diminished $T_{\text{failover}}$ value signifies enhanced failover efficiency. This is essential for reducing disturbance during server outages.

\section{Results and Discussion}\label{sec4}

\begin{figure}
    \centering
    \includegraphics[width=0.80\linewidth]{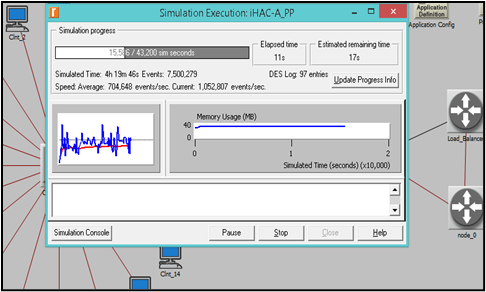}
    \caption{Executing simulation}
    \label{fig:fig12}
\end{figure}

\begin{figure}
    \centering
    \includegraphics[width=0.90\linewidth]{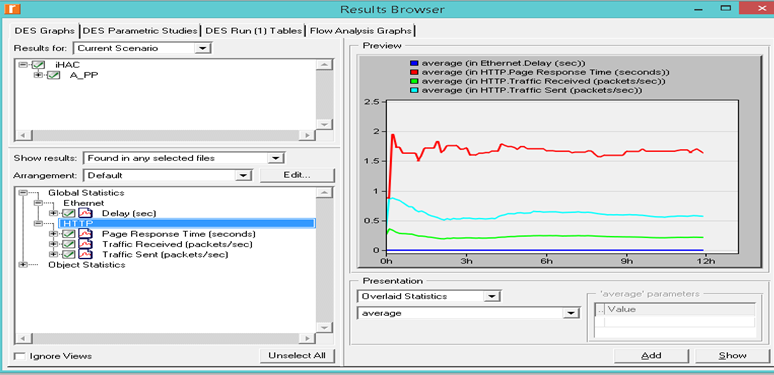}
    \caption{Results of the simulation}
    \label{fig:fig13}
\end{figure}

The Integrated High Availability Cluster (iHAC) was evaluated against active-active and active-passive cluster where we used Riverbed Modeler to show the efficacy in tackling critical difficulties related to high-availability (HA) systems. The tests evaluated iHAC effectiveness across various configurations, demonstrating its utility during hardware or system malfunctions. Results highlight the capacity to sustain high availability by reducing latency, enhancing throughput, and guaranteeing operational continuity during interruptions. The Riverbed Modeler uses a discrete event-driven simulation (DES) which facilitates accurate network performance and behavior evaluation. Figures \ref{fig:fig12} and \ref{fig:fig13} depict the simulation execution procedure and its outcomes. To replicate near-realistic scenarios, the simulation time was limited to 12 hours. In about 4 hours, the system reached an average speed of 704,648 events/sec, resulting in an HTTP traffic peak. The configuration decision optimizes computational efficiency while ensuring a realistic representation, given that data center servers’ function constantly over prolonged durations.

\subsection{Performance Analysis: Active-Active Configuration}\label{subsec4}

\begin{figure}[h]
    \centering
    \includegraphics[width=1\linewidth]{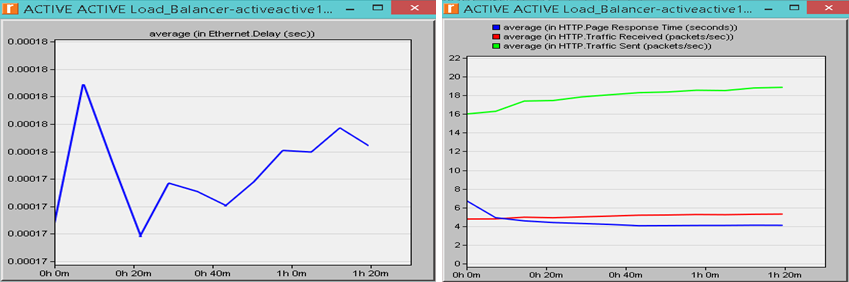}
    \caption{Active-Active Cluster Ethernet Delay and Simulation Result}
    \label{fig:fig14}
\end{figure}

 The simulation result for the active-active cluster configuration is presented in Figure \ref{fig:fig14}. Delay statistics reveal minimal variation, starting at 0.00017 and peaking at 0.00018 before stabilizing. This pattern is attributed to the load distribution across servers, where all deployed servers actively share client requests. The traffic sent and received, as well as the HTTP page response times, exhibit consistency. Traffic sent starts at approximately 16 packets per second and gradually increases, while traffic received begins at around five packets per second and rises incrementally. The HTTP response time stabilizes at an average of five seconds due to the Least Connection algorithm employed in the load balancer, ensuring equitable resource allocation among servers.

\subsection{Performance Analysis: Active-Passive Configuration}\label{subsec4}

\begin{figure}[h]
    \centering
    \includegraphics[width=1\linewidth]{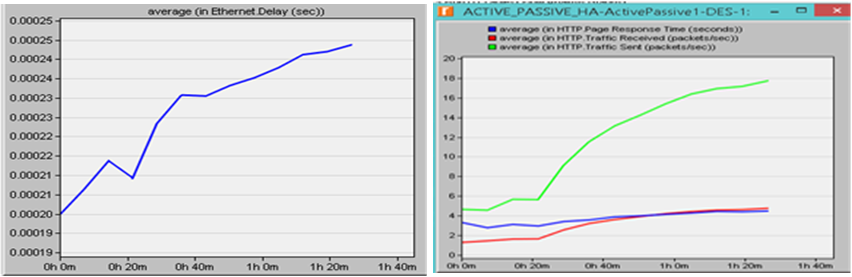}
    \caption{Active-Passive Cluster Ethernet Delay and Simulation}
    \label{fig:fig15}
\end{figure}

Figure \ref{fig:fig15} illustrates the active-passive cluster and results. The Ethernet delay begins at 0.00020 and fluctuates as server one operates actively and handles all client requests. This configuration results in an increased latency rate compared to the active-active scenario, where multiple servers handle requests simultaneously. When server one fails, server two transitions into the active state, ensuring service continuity. Despite the increased delay, this setup may maintain business operations during server failures, but it is not guaranteed.

\subsection{iHAC Generic Model Performance}\label{subsec4}

\begin{figure}
    \centering
    \includegraphics[width=1\linewidth]{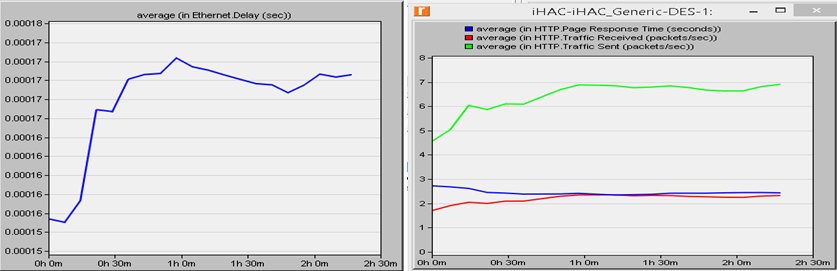}
    \caption{iHAC Generic Model Simulation Result}
    \label{fig:fig16}
\end{figure}

Figure \ref{fig:fig16} depicts the simulation results for the iHAC generic model, featuring two host servers, each running two virtual machines. Traffic entering the system is categorized as HTTP heavy or light browsing, with the primary server handling both traffic types in an active state. Passive servers support the primary server by dedicating resources to light browsing while maintaining readiness to transition into an active state when there is a failure. The simulation statistics for the iHAC model indicate a delay beginning at 0.00016 and peaking at 0.00017, stabilizing thereafter. On average, seven packets are sent per second, while traffic received fluctuates between two and three packets per second. The HTTP page response time averages three seconds under normal operations, showcasing the iHAC ability to balance the load and effectively reduce response times.

\subsection{Comparative Insights and iHAC Strengths}\label{subsec4}

The experiments demonstrated that iHAC surpasses conventional designs by enhancing resource usage and reducing delays. It performs better in all circumstances, with diminished Ethernet latency and accelerated HTTP response times. The active-active configuration guarantees equitable load allocation. This results in consistent performance measurements. The active-passive technique emphasizes failover reliability by ensuring uninterrupted operation in the event of server failures. The iHAC generic model combines the strengths of both traditional architectures (active-active and active-passive), providing robust flexibility and efficiency. The simulation results effectively confirm iHAC’s ability to address latency, throughput, and availability. An average HTTP response time of three (3) seconds in the iHAC model enhances user experience and operational reliability, making it an ideal choice for system resilience.

\section{Conclusion and Future Work}\label{sec5}

This paper introduced and evaluated the Integrated High Availability Cluster (iHAC), a hybrid architecture designed to address the inherent trade-offs of traditional active-active and active-passive clusters. Our simulation results demonstrate that by intelligently combining the principles of both legacy architectures, the iHAC architecture achieved superior performance, significantly reducing latency and HTTP response times while optimizing resource utilization. The scalable design, which leverages advanced load-balancing algorithms and virtualization, provides a practical and effective solution for enhancing system resilience against hardware and system failures.
While this study successfully demonstrates the performance benefits of the iHAC, it is important to acknowledge its limitations, which in turn provide clear directions for future research. Our evaluation focused on key performance metrics such as response time and throughput under specific failure scenarios. Other metrics, such as packet loss rates during failover events, were not a primary focus and represent a valuable avenue for future work to more deeply quantify the seamlessness of the recovery process. Furthermore, while the 12-hour simulation period provided stable results, longer-term testing under more diverse network conditions would be beneficial to identify any potential bottlenecks or performance patterns not observed in this initial evaluation. Future research could also explore the efficacy of more advanced load-balancing techniques or the use of "magic packets" to further optimize the wake-up time for passive servers, thereby continuing to enhance the scalability and robustness of the iHAC architecture.

\sloppy
\bibliography{sn-bibliography}

\end{document}